\documentstyle[12pt]{article}
\include{epsf}
 
\newlength{\extralineskip}
\newcommand{\la}{\langle}
\newcommand{\ra}{\rangle}
\newcommand{\beq}{\begin{equation}}
\newcommand{\eeq}{\end{equation}}
\newcommand{\bd}{\begin{displaymath}}
\newcommand{\ed}{\end{displaymath}}
\def\tr{{\rm tr}}
\def\e{~{\rm e}}
\def\bea{\begin{eqnarray}}
\def\eea{\end{eqnarray}}

\def\ba{\beq\new\begin{array}{c}}
\def\ea{\end{array}\eeq}

\def\lie{{\cal L}}

\def\inbar{\,\vrule height1.5ex width.4pt depth0pt}
\def\IC{\relax\hbox{$\inbar\kern-.3em{\rm C}$}}
\def\IR{\relax{\rm I\kern-.18em R}}
\def\IZ{{{\rm Z}\!\!{\rm Z}}}
\makeatletter
\newdimen\normalarrayskip              % skip between lines
\newdimen\minarrayskip                 % minimal skip between lines
\normalarrayskip\baselineskip
\minarrayskip\jot
\newif\ifold             \oldtrue            \def\new{\oldfalse}
\def\arraymode{\ifold\relax\else\displaystyle\fi} % mode of array entries
\def\@arrayskip{\ifold\baselineskip\z@\lineskip\z@
     \else
     \baselineskip\minarrayskip\lineskip2\minarrayskip\fi}
\def\@arrayclassz{\ifcase \@lastchclass \@acolampacol \or
\@ampacol \or \or \or \@addamp \or
   \@acolampacol \or \@firstampfalse \@acol \fi
\edef\@preamble{\@preamble
  \ifcase \@chnum
     \hfil$\relax\arraymode\@sharp$\hfil
     \or $\relax\arraymode\@sharp$\hfil
     \or \hfil$\relax\arraymode\@sharp$\fi}}
\def\@array[#1]#2{\setbox\@arstrutbox=\hbox{\vrule
     height\arraystretch \ht\strutbox
     depth\arraystretch \dp\strutbox
     width\z@}\@mkpream{#2}\edef\@preamble{\halign \noexpand\@halignto
\bgroup \tabskip\z@ \@arstrut \@preamble \tabskip\z@ \cr}%
\let\@startpbox\@@startpbox \let\@endpbox\@@endpbox
  \if #1t\vtop \else \if#1b\vbox \else \vcenter \fi\fi
  \bgroup \let\par\relax
  \let\@sharp##\let\protect\relax
  \@arrayskip\@preamble}

\begin{document}

\begin{titlepage}
\setcounter{footnote}0
\rightline{\baselineskip=12pt\vbox{\halign{&#\hfil\cr&
hep-ph/9701270 &\cr
{   }&\cr 
%&Revised Version {   }&\cr &\today
\cr}}}
\vspace{0.5in}
\begin{center}
{\Large\bf Vacuum Structure of
Two-Dimensional Gauge Theories for Arbitrary Lie Groups}\\
\medskip
\vskip0.5in
\baselineskip=12pt

\normalsize {\bf L. D. Paniak}\footnote{E-mail: paniak@physics.ubc.ca},
{\bf G. W. Semenoff}\footnote{E-mail: semenoff@physics.ubc.ca}
and {\bf A. R. Zhitnitsky}\footnote{E-mail: arz@physics.ubc.ca}
\medskip

\baselineskip=12pt
{\it Department of Physics and Astronomy, University of
 British Columbia\\
Vancouver, British Columbia, Canada V6T 1Z1}
\end{center}
\vskip1.5in
    
\begin{abstract} \baselineskip=12pt 
Using the well established
machinery of Wilson loop calculations we investigate the multiple
vacua of two dimensional Yang-Mills theories with infinitely massive
adjoint matter.  In particular, via group theoretical techniques
we calculate string tensions between charges and find the number of
vacua for each compact Lie symmetry group.  
The counting of vacua is in agreement with the standard
classification based on the topology of the effective gauge group
$\pi_1(G/Z)$ when one considers arbitrary numbers of adjoint charges
in the system.  For systems with limited numbers of charges we find
additional "meta-stable" vacuum states.  Finally we discuss t'Hooft's
disorder operators in this setting as number operators for the
multiple vacua.  
\end{abstract}

\end{titlepage}
\section{Introduction and Motivation }

After the invention of instantons \cite{polyakov} it became clear 
that the vacuum structure of gauge theories is non-trivial and 
has definite effects on the physics of the theory.
The existence of large gauge transformations 
and consequent non-contractable paths in the space of fields leads
to a parameterization of distinct vacuum states by the  
periodic vacuum angle $\theta$ (\cite{Gribov}-
\cite{CDG}). Unfortunately, this simple and
apparently universal picture is incomplete as shown by the 
well known four-dimensional examples of 
gluodynamics with large $(N_c)$ numbers of colours, supersymmetric
QCD with arbitrary numbers of $(N_f)$ flavours and colours and
QCD with massless quarks, in addition to others.
In all these cases we find that the parameter $\theta$ does not
appear simply as the argument of a periodic function as a standard classification based on integer topological numbers would predict.
Rather, $\theta$ dependence in the theory comes about through
the ratio $\theta/N$ which seems to suggest a new classification
principle.  Indeed,
in the chiral $(m=0)$ limit of four-dimensional QCD, Ward identities~\cite{Crewther}
imply $\theta$ dependence appears only 
through the ratio $\theta/N_f$ and, for a fixed value of $\theta$, 
there are still $N_f$ different vacuum states.  
Similarly in gluodynamics \cite{WV}, $\theta$ dependence appears 
as $\theta/N_c$  
\footnote{We recall that the
$\theta $ dependence of physics is linked to the $U(1)$ problem.
Indeed, if we believe that the resolution of $U(1)$ problem appears
within the framework of the papers \cite{WV}, we must assume that the
topological susceptibility $$K=i\int d^4x\la
0|T\{Q(x),Q(0)\}|0\ra\sim \frac{d\la 0|Q |0\ra}{d\theta}\sim
\frac{1}{N_c}$$ in pure gluodynamics is not zero and
$K\sim\frac{1}{N_c}$.  Here
$Q=\frac{1}{32\pi^2}G_{\mu\nu}\tilde{G_{\mu\nu}}$ is topological
density.  It demonstrates one more time that the dependence on
$\theta$ comes through $\theta/ N_c$. In particular, it is expected
that $\la Q\ra\sim\sin(\frac{\theta}{N})$}.  
Consequently we can
retain $2 \pi$ periodicity in the variable $\theta$ only if
the vacuum is $N_f$ or $N_c$-fold degenerate, respectively
for each value of $\theta$.
Labeling these degenerate vacua implies an additional
superselection rule over and above the original $\theta$
parameterization.  In an important test of this picture of vacuum
structure, we will investigate the nature of this 
superselection rule in a solvable model.
 
In addition to gaining direct results about the details of 
vacuum structure, we will be considering a solvable model
in part due to inconsistencies amongst different techniques for
investigating the vacuum structure of gauge groups. 
For example, consideration of the Witten
index and other aspects of supersymmetric theories in four dimensions
gives contradictory results in these cases.
More precisely, in four dimensional supersymmetric field theories with
arbitrary gauge group, the exactly calculable gluino condensate
$\la\lambda\lambda\ra \sim \e^{i \theta/N}$ 
is not vanishing \cite{SV}.  This result is
unexpected since the number of fermionic zero modes in the tunneling
transition associated with the standard instanton is larger than two
for any gauge group. Consequently, this analysis predicts that the
number of vacuum states is equal to the quadratic Casimir of adjoint
representation $C_2(Ad)$.  In general this is in conflict with an
analysis of the Witten index \cite{witten2} for supersymmetric QCD.
For a gauge group of rank $r$ the Witten index 
predicts  the number of vacua to be $r+1$ (independent
of space-time dimension) which differs from the quadratic Casimir
except in the cases of $SU(N)$ and $Sp(N)$.

Additionally, contradictory results about vacuum structure are not limited to four dimensions as shown in the analysis of 
two dimensional QCD with dynamical fermions in adjoint representation,
(\cite{witten}-\cite{kogan}).  In particular, there is the following
discrepancy: Standard bosonization techniques (and large $N$ arguments
\cite{kogan}) predict the existence of a fermion condensate for an
arbitrary gauge group.  At the same time, the standard topological
classification of vacuum states with the corresponding counting of
zero modes, does not seem to support such a condensation. Such a
discrepancy in the two dimensional theory is very similar to what we
discussed above in four dimensional supersymmetric models where there
existence of a gluino condensate does not appear to be supported by
vacuum transitions with the standard properties.

In short, it is clear that current understanding of the 
classification of vacuum states is not complete.
More sophisticated methods of classification may be
necessary in order to match well
known (but indirect) results with the analysis of admissible large
gauge transformation in gauge theories. 
Before increasing the complexity of the analysis though,
it would be very useful to
understand the problems formulated above on a more detailed level in
toy models.  Thus our {\bf main goal} is an analysis of the
multiplicity of vacuum states in two dimensional gluodynamics with
arbitrary (compact Lie) gauge group.  Recently the problem of
classification of vacuum states for the model based on an $SU(N)$
gauge group has been discussed \cite{psz} (see also \cite{sodano} for
a classification in the finite temperature case). There it was
explicitly shown that the model exhibits exactly $N$ different vacuum
states.  In addition, the string tension and vacuum energy in each 
given vacuum state were found using two independent approaches-
the standard Hamiltonian approach and Wilson loop calculations.
Using the same machinery here we will see that 
all alternative vacua corresponding to the extra superselection
rule follow from a topological classification of the large
gauge transformations. We emphasize that our results are strictly 
from group theoretic calculations to determine the number of stable
ground states and do not involve direct input from topology.

The paper is organized as follows.  In the 
next section we will outline the topological classification 
of vacua in gauge theories and the connection to boundary 
conditions. Once one has particular boundary conditions
then the classification problem can be recast in terms of
Wilson loops and, for two-dimensional gauge theories, group
theory. In discussing some of the details of the group theoretical
calculations, it will become clear that counting true vacua
with non-unitary gauge groups is somewhat more complicated
than the previously dealt with unitary case.  In particular
we will demonstrate the existence of `meta-stable' vacua 
in the (unphysical) limit when the charges in the system are
infinitely massive.  Finally, we will consider the problem of
enumerating vacua in two-dimensional gauge theories from a dual
perspective in terms of disorder operators.

\section{Topology and Boundary Conditions}

The standard method of classifying the multiplicity of vacua
in a particular gauge theory with adjoint matter hinges on
identifying the effective gauge group. Here, since gauge
transformations operate by adjoint action on all fields,
the true gauge group is the quotient of the gauge group and its
center.  This quotient is multiply connected.  For simply
connected
semi-simple gauge group $G$ with center $Z$,
\beq
\pi_1( G/Z)~=~ \pi_0(Z)~=~ Z
\eeq
This gives a classification of gauge fields which are
constrained to
be flat connections at infinity.  In that case
\beq
\lim_{ |x| \rightarrow \infty}~A_\mu(x)~=
ig^{\dagger}(x)\nabla_\mu
g(x)
\eeq
Where $g(x)$ is a mapping of the circle at infinity to the gauge group
$G/Z$. Since $G/Z$ is a symmetry of the Hamiltonian,
we expect that all physical states carry a representation
of $\pi_1(G/Z)$. In the case where the center of the group is Abelian
all of its irreducible representations are one dimensional
and further, when $Z \sim \IZ_{n_1} \times \cdots \times
\IZ_{n_j} $,   we are
lead to a classification of all physical states in terms of 
$j$ generators of $Z$, $\{z_1, \cdots, z_j \}$.
 If ${\cal Z }$ is a unitary realization of $Z$ and 
$| \psi >$ is a physical state we have
\beq
{\cal Z} |\psi > = \e^{ i (z_1 + \cdots +z_j)} | \psi >
\eeq

A more direct way to classify vacua in the case of SU(N)
gauge group, without resorting to topological arguments,
was developed in \cite{psz} following the example
of multiple vacua in the massive Schwinger model (\cite{coleman}, 
\cite{suss}). These methods were previously introduced
by Witten \cite{witten} to identify the existence of
$\theta$-vacua in two dimensional non-Abelian field theories.
To review, in the Schwinger model multiple vacua are
considered as generalized boundary conditions on the model where
static charges reside at either end of the spatial dimension of
the world. In the non-Abelian case we have the same situation
except that the c-number charges are replaced by static Lie
algebra valued colour charges $T_R$ and $T_{\bar{R} }$ at 
either end of the world. Here the $T$'s are the generators
of the colour group in the representation $R$
and its conjugate, respectively and form a discrete choice of
boundary conditions as opposed to the continuum in the
Schwinger model.

The key to utilizing such a picture of multiple vacua
as generalized boundary conditions hinges on the fact that static
charges in a gauge theory have a natural interpretation  in 
terms of Wilson loops as can be seen directly in the
Hamiltonian formulation of the problem \cite{psz}.
If $\tr_R$ is the trace in the 
representation $R$ of the gauge group then we can impose
boundary conditions directly in the action by including a
Wilson loop
\beq
Z \rightarrow \int {\cal D} \psi {\cal D} \bar{\psi} {\cal D}
 A
\exp{(- \int d^2 x ~\lie )}~
\tr_R P \exp{(i \oint_{C \rightarrow \infty} dx^\mu A_\mu (x) )}
\label{bloop}
\eeq
The number of distinct choices of representation in which
to take the trace of the boundary loop is directly related to the
transformation properties of the loops under $Z$, the center of
the gauge group due to the prescence of only adjoint charges in
the system.  Since different boundary charges may be equivalent
up to decays via adjoint charges, there are only a
small number of stable vacua in any theory with dynamical
charges since all decays are possible in this case.
What is interesting though is that the number of stable
vacuum states for a theory with static charges depends on the 
number of charges in the system and is in general greater or equal
to the number of stable vacuum states in the dynamical theory.  
This contrasts the case of SU(N) which was considered
previously where there are $N$ stable vacuum states for both
static and dynamical matter. Investigating such differences
was a main motivation for the present work since,
as mentioned earlier, in four dimensional theories
a similar discrepancy is observed for the gauge groups
other than $SU(N)$.

\section{Calculations with Wilson Loops}
First of all let us recall
why the calculation of nested Wilson loops is an appropriate
method for analysis of the vacuum structure of the theory.
As is known,  different vacuum states are classified
by different boundary conditions which can be expressed
by formula (\ref{bloop}) with a Wilson loop inserted at infinity.
At the same time if we are interested in the 
string tension for a very heavy charge-anti-charge pair
of a particular representation R in a given vacuum state,
this is nothing but the calculation of a loop in representation R
in the background of the external loop and similarly for additional
charges.
The vacuum expectation value of corresponding system can be calculated by using the standard technique (\cite{mak}-
\cite{gromat}) and the string tension can be easily extracted.
What is important for us  is the observation that a stable vacuum
state corresponds to the situation when  string tension for all
states is a {\bf non-negative number}. Negative string tension between
a pair of charges corresponds to the unstable situation where it
is energetically favourable for the charges to be infinitely separated. 

Let us review the 
previous calculation \cite{psz}. The configuration we are interested
in here is that of a pair of adjoint charges in a fundmental
background with the Wilson correlator of a single
loop in the adjoint representation of SU(N) with boundary $C_2$ 
enclosing total area $S_2$ nested within a loop $C_1$. The contour
$C_1$, which encloses an area $S_1+S_2$, is taken in the 
k-fold anti-symmetric (fundamental) representation $(k=0 \ldots N-1)$.
The expectation value $<W(C_1,C_2)>$ of such a configuration can be calculated explicitly with the result
\bea
<W(C_1,C_2)>
=&\frac{ N!}{(N-k)! k!} \e^{-g^2(S_1 + S_2)
\frac{k(N-k)(N+1)}{2N}}
[ 1 + \frac{(N-k-1)(N+1)}{k+1} \e^{ -g^2 S_2 (N-k) }
\label{sucase} \\
&+ \frac{ (N+1)(k-1)}{N-k+1} \e^{-g^2 S_2 k}
+\frac{k N(N+2)(N-k)}{(k+1)(N-k+1)} \e^{-g^2 S_2 (N+1)}] \nonumber
\eea
Where we will take the leading factor, which is just the contribution of a k-fundamental loop of total area $S_1+S_2$, to be normalization which
physically corresponds to a constant colour-electric field throughout
the one dimensional world.
The relevant part of the formula, in square brackets, describes  
different states with string tensions $0$, $g^2 (N-k)$, 
$g^2 k$ and $g^2 (N+1)$. Since for each $(k=0 \ldots N-1)$ these
are non-negative we conclude that for a theory based on an SU(N)
gauge group the number of vacuum states is equal to $N$.
If we had considered different (non-fundamental) representations
for the external loop which labels vacua,
we would have found negative string tensions for some states-those
corresponding to unstable charge configurations.
Of course, this result was anticipated from more general
considerations \cite{witten}, but we believe that the
explicit demonstration of instability is an important
check of our formalism.

With this example in mind we are ready precisely formulate our
approach to enumerating the vacuum states of 1+1 dimensional Yang-Mills
theories with heavy adjoint matter.
As in the unitary case, we will calculate the vacuum expectation value
of the system of  adjoint Wilson loops bounded by a single
loop taken in different representations.
Only those boundary representations which lead to non-negative
string tensions count as vacuum states otherwise the configuration 
is unstable and should be excluded from the consideration.
Looking at systems with more than one pair of adjoint charges
(more than one adjoint loop) it will become clear that stability
is a more complex issue for a general gauge group than for SU(N).
These complications will lead us to consider generalizations of
the notion of N-ality for arbitrary Lie and
a complete classification of vacua.

\section{Group Theory and Counting Vacua}

In this section we present some explicit calculations of the
enumeration of vacuum states for arbitrary compact Lie groups.
We should note that these calculations are completely independent
of the topological considerations of Section 2 as we will use group theoretic methods to calculate string tensions directly in 
different vacua. Of course, anticipating our results, we note
that the calculations of this sections are in general agreement
with the topological classification.
In terms of group theory this enumeration
involves calculation of the Kronecker products of arbitrary representations with that of the adjoint and 
decomposing the result into a sum of irreducible representations.
While it is difficult to show in general
that a non-fundamental representation as a boundary charge leads to an unstable
configuration, it will become clear later that we need only consider fundamental
external charges and hence we will only consider these.  To begin we will consider an example of 
the symplectic groups Sp(2N) since they are similar in complexity to the unitary
group and they show the basic characteristics of the considerations we must make for general simple Lie groups.

In Sp(8) each of the representations is given by a Young table, a tensor 
representation in contrast to the spinor representations which arise in the
orthogonal groups.  As Sp(8) is 36 dimensional we can easily identify the adjoint
representation with the symmetric combination of two 8 dimensional fundamental
representations; $(2000)$ in Young table notation as detailed in the appendix.
The other three fundamentals are given by  anti-symmetric
combinations of the 8 dimensional $(1000)$ representation: $(1100)$, $(1110)$
and $(1111)$ which are 27, 48 and 42 dimensional 
respectively. From tables of branching rules  
\cite{branch} or directly from a table of products \cite{butwyb} one can 
easily find the product of each of the fundamentals with an adjoint representation.
Using the expressions in the appendix for the quadratic Casimirs of arbitrary
tensor representation for the classical Lie groups and subtracting the 
Casimir of the external fundamental we have the string tensions for 
states occurring in each loop configuration
\beq
36 \otimes 8: ~ 0,\frac{9}{2} g^2,\frac{11}{2} g^2 ~~~~~~~
36 \otimes 27:~ 0, g^2,5 g^2,6 g^2
\label{sp8tensions}
\eeq
\bd
36 \otimes 48:  0,\frac{3}{2} g^2,\frac{7}{2} g^2,6 g^2~~~~~~~
36 \otimes 42:  0,3 g^2,6 g^2
\ed
It appears that each fundamental vacuum is stable
under interaction with a single adjoint. Explicit calculations for other symplectic
groups suggest that this is the general case with the number of stable vacua
for Sp(2N) equal to N+1.
It is interesting to compare this result with naive expectations
from four dimensional calculations mentioned in the introduction.
Here, N+1 different vacuum states agrees with a classification
based on the adjoint quadratic Casimir as in supersymmetric theories \cite{SV},
and calculations of the Witten index but clearly differs from 
general topological
considerations where the vacua should be a representation of 
$\pi_1(G/Z)\sim \IZ_2$ with only two elements. 

The other classical Lie groups we have yet to consider are the orthogonal
groups.  Typically these groups are more complicated to deal with 
since they have spinor representations which cannot be characterized 
by a Young tableaux.  For our purposes though the presence of spinor 
representations simplifies matters in the following way.  Since the 
adjoint representation of an orthogonal group SO(N) is given by the 
anti-symmetric combination of two N-dimensional vector representations,
the adjoint is a tensor representation and cannot interact with a spinor
to produce an unstable configuration.  In short we need only consider the 
stability of tensor fundamentals interacting with  adjoint charges.
Since we are considering only pure tensor representations, the arguments
for SO(2N+1) also apply to SO(2N). With this in mind we consider in detail
only the case of SO(9).

For SO(9) the adjoint is 36 dimensional and represented by a Young table
$(1100)$.  Since it also a fundamental there are only two tensor 
fundamentals to consider in this case.  First is the 9 dimensional 
vector representation $(1000)$. The normalized string tensions for 
the configuration of a single adjoint loop with this external 
charge are $0, 5 g^2 , 16 g^2 $ and we see that all these 
configurations are stable.  The other tensor fundamental is 84 dimensional
with Young table $(1110)$.  Here we begin to see some differences with 
the case of SU(N) in that this fundamental does {\bf not}  
form a stable configuration with a single adjoint loop.  This is 
clear from the normalized string tensions $-5 g^2$, $0$, $ g^2$, $3 g^2$,
$6 g^2$ and $9 g^2$. Consequently we find for the case of SO(9) there
are three different stable vacuum states corresponding to the trivial
vacuum, one with a single spinor representation as a boundary charge and
one with the 9 dimensional vector representation as a boundary charge.
This situation is in fact generic.  For the orthogonal groups one can 
show using branching rules (\cite{hammer}, \cite{butwyb})
that all tensor
fundamentals consisting of anti-symmetric combinations of more than 
two vector representations are unstable when combined with adjoint
charges.  Hence for SO(2N+1) we have three stable vacuum states and 
for SO(2N) there are four due to the second spinor representation.
Remarkably these results disagree with naive expectations of 
supersymmetric calculations,
the Witten index and  topological arguments.

Next we will investigate the five exceptional Lie groups. 
We will consider only the simplest, $G_2$ of rank two in any detail.
Consulting standard group theory references \cite{groupth} we find that the first fundamental is 7 dimensional and the second, which is also the adjoint,
is 14 dimensional.
Consequently there are only two different possible 
vacuum states here: the trivial and the one given by a 7 dimensional
fundamental boundary charge which calculation \cite{butwyb}
shows to be stable acting with an adjoint. 
Again we find that this result agrees with 
the supersymmetric calculation where the number of vacua is given by 
the quadratic Casimir of the adjoint representation but disagrees with 
the Witten index calculation and topology.  By homotopy arguments 
since $G_2$ is simply connected and it has a trivial center we have trivial
$\pi_1(G/Z)$ and expect only a single, trivial vacuum.

In a similar fashion one can investigate the stability of fundamental
boundary charges for  the other exceptional 
Lie groups  $F_4$, $E_6$, $E_7$ and $E_8$.  Using Kronencker product tables
from \cite{slansky} one finds that for $F_4$ only the 26 dimensional
fundamental gives a stable configuration when combined with a single 
adjoint. For $E_6$ there are two degenerate 27 dimensional fundamentals
each of which are stable and $E_7$ has only a single 56 dimensional 
fundamental which results in a stable configuration.  The last of
the exceptional groups, $E_8$ has no fundamental representations leading
to stable configurations. This exhausts all of the possible 
cases for simple compact Lie groups and the results that we have obtained
are listed in Table 1.  The results of this section are
suggestively place in the row labeled meta-stable vacua and comparing
with the group theory and topological information tabulated there it
is clear that none of the classification schemes we have been considering
are accurate.
Naively, one would think that
all the vacua we have  demonstrated here are perfectly stable
with respect to the  creation of a pair of adjoint charges
and should be taken into account.
At least, in case of $SU(N)$  such a procedure reproduced all of the 
know results for enumerating the multiple vacua of the theory.
Clearly an understanding of what is going on for SU(N) theory is
not sufficient to understand the general situation.  This is obvious
if one notes that the different approaches we have been 
considering: gluino condensates, the Witten index and the topology of the 
configuration space all give the same number of vacua for SU(N) - N.
The essential details of the situation are not clear from the unitary
case and we will need to investigate the general case more carefully.

Up to this point we have been considering  the stability problem 
only with respect to creation of a pair of adjoint particles. 
This approach is perfectly general in Abelian gauge theories 
where the interactions have the same character for
arbitrary numbers of interacting charges.
However there are complications in the case
of non-Abelian theories where interaction between different
representations become quite involved. Before analyzing 
this more general case
we return momentarily to the example of $SU(N)$.
Consider multiple adjoint loops in the standard
background of a fundamental loop which corresponds
to the possibility of creation of arbitrary numbers of particles.
In SU(N) the adjoint by definition transforms trivially under the center 
(has no N-ality) and each of the fundamentals transforms with a different phase 
(has a different N-ality) so each is stable \cite{witten} under decays via adjoint charges into each other or the trivial representation.
What is important for us is that this statement does not depend
on the number of internal loops. The crucial element of this observation is 
the existence of a conserved quantum number: N-ality. Identifying  the 
analogous feature of a general Lie group will lead us to the stable vacuum states.  

Consequently we need to check is if a boundary charge is stable against decay via multiple  adjoint charges.  For SU(N) this was not necessary due to the different  N-ality of the fundamentals but in the general case we will
use stability to determine the analog of N-ality for a general Lie group.
As a concrete example let us return to the case of Sp(8) for 
which we have seen four different, apparently stable vacua (\ref{sp8tensions}).
Instead of a single adjoint loop we now consider the system containing
two adjoint Wilson loops nested within a single fundamental loop.
Consider for instance the interaction of two adjoint 
representations with
a single 27 dimensional fundamental: $36 \otimes 36 \otimes 27$.
In the process of calculating  (\ref{sp8tensions}) one finds the 
decomposition of the Kronecker product
\beq
36 \otimes 27 = 27 \oplus 36 \oplus 315 \oplus 594
\eeq
and hence adding a second adjoint representation to the product produces 
a term $36 \otimes 36$ which certainly contains the trivial representation
and will correspond to states with negative string tension once one 
renormalizes by the 27 dimensional boundary loop. By considering a similar
process with more adjoint factors, the 42 and 48 dimensional representations
can be shown to be unstable as well. Consequently we find that while 
all of the fundamental charges of Sp(8) are stable against decay via 
one adjoint charge, only the 8 dimensional one appears stable against
an arbitrary number. While one cannot test the general situation for 
arbitrary groups via the method of loop calculations there is a different approach which will lead us to the correct 
generalization of N-ality for general Lie groups.

By now it is clear that  the stable fundamentals which define 
different vacua for a given group are exactly the
fundamental charges which cannot decay into each other nor the trivial
representation via interactions with adjoint charges.  In the physical 
picture these mark the end of decay chains and are the minimum energy 
representatives of their respective classes.  Mathematically this idea
is formalized in the concept of minimal representations which 
have been utilized before in the physics literature especially by
Goddard and Olive \cite{golive} in discussions of monopole stability. 
As noted there, if we have the lattice of weights
$\Lambda(G)$ for a simply connected compact Lie group $G$ then the lattice
of roots (weights of the adjoint representation) $\Lambda_{Ad}(G)$  is a subgroup of it.  Consequently we can form the cosets $\Lambda/ \Lambda_{Ad}$
by identifying weights which differ by integral linear combinations of 
roots and in this way we identify precisely which representations lie in like
decay chains.  These cosets form a finite dimensional Abelian group  which
is isomorphic to the center $Z$ of the group $G$ and now the importance
of the center in classifying vacuum states becomes obvious.
Clearly the notion of
N-ality for the unitary groups is generalized here where to each coset 
one can assign a (conserved) element of an Abelian group- typically, but
not always, an integer.

Once we have determined the classes to which different representations 
belong the question remains: Is there a representative of each class
which is stable?  The answer to this question is yes.  First we deal with
the identity coset where the representation with minimum quadratic Casimir
(energy) is just the trivial representation.  Clearly this coset includes the 
adjoint representation and all other representations which transform 
trivially under the center of the group.  As for the other cosets we can 
equivalently define the minimal weights $\theta$
as those which lie closest to the
origin of the weight diagram or have the least Casimir or where all
weights $\{ \theta \}$ of a representation lie on the same orbit under
the action of the Weyl group.  The positive (dominant) elements of
the minimal weights then identify the minimal representations of the group
and identify the complete set of stable charges which label
non-trivial vacuum states for any Lie group. 

It is now an easy task to enumerate the minimal representations 
for all simple, compact Lie groups. In the case of the
special unitary group SU(N) these are of course the N-1 fundamental
representations. For the symplectic
group Sp(2N) only the 2N dimensional fundamental is stable in interactions
with adjoint charges.  The other classical Lie group is the orthogonal 
group SO(N) and here we have two different cases. For SO(2N+1) only the
$2^N$ dimensional spinor representation is minimal but for SO(2N) things
are more interesting where both of the spinor representations are stable
in addition to the N dimensional vector representation. As for the 
exceptional groups, both 27 dimensional fundamentals of $E_6$ are 
minimal as is the 56 dimensional fundamental of $E_7$.  The remaining
compact Lie groups $E_8$, $F_4$ and $G_2$ have no non-trivial minimal 
representations, that is all representations can decay to the trivial
one via adjoint charges.  Now that we have identified all possible 
boundary charges which give rise to non-trivial vacua we see that all
candidates are fundamental representations justifying a restriction
we placed on ourselves earlier.  We now tabulate this information
in the last two rows of Table 1.  The number of stable 
vacua is just the number of minimal representations in addition
to the trivial representation which labels the standard perturbative
vacuum.
 
Some comments are in order.  We have shown that when enumerating the 
stable vacuum states of Yang-Mills theories with heavy adjoint matter the 
center symmetry of the  group is crucial.  Neither supersymmetric
nor Witten index calculations lead to consistent results but topological
considerations, namely the connectivity of the effective gauge group
$G/Z$ does.  In particular, the group structure of the 
center gives the structure of the vacuum states and in particular
if representations $R_1$ and $R_2$ label different stable vacua
then the multiplication table of the center determines the character
of the vacuum with label $R_1 \otimes R_2$.  

The second remark is much more speculative: We found that in
the limit of very heavy   quark mass, the meta-stable
vacuum states will be stable with respect to   decays by 
a single pair of adjoint charges  but not in more complicated scenarios.
For instance, in the chiral limit
when an arbitrary number of massless quarks may appear, those vacuum states
are certainly unstable. Since the number of stable and 
meta-stable vacua are different for a given theory 
one could expect a phase transition with variation of  the quark mass
for those $\theta$ vacuum states which are meta-stable, rather than stable.
If this indeed true, such a transition might have some relation
to the problem of   chiral condensation in $QCD_2$
with adjoint matter we discussed in the introduction.

\section{Disorder Dual Operator }

Now we turn to a discussion of confinement phenomena in two-dimensional
gluodynamics from a point of view dual to the one we have been 
using to enumerate vacuum states.  Of course, as we have shown in the
previous sections, all physics is determined by the effective 
gauge group $G/Z$ so a dual picture has little new to offer.  Regardless, we
feel that a different point of view on the same phenomenon
is desirable and in this section we investigate these confinement
properties in terms of a so-called disorder operator $M$. As we
will show, $M$ leads to algebraic structures analogous to
number and ladder operators for multiple vacua in a gauge theory.

The importance of disorder operator in gauge theories has been
emphasized by 't Hooft\cite{Hoof1} (see also \cite{Pol}),
who argued that rather than instantons it is the field configurations with nontrivial $\IZ_N$ topological properties (for $SU(N)$ gauge group) that should be considered responsible for long range interactions. Analogous disorder
variables have been used in different fields of physics (\cite{Kad}-\cite{Zam}) and most of the ideas described here have been extracted
from the classical papers (\cite{Hoof1}-\cite{Zam}). 
In what follows we deal with the specific case of an $SU(N)$ gauge
group however generalization to any compact Lie group follows by  
replacing $\IZ_N$ factors by elements from the center of the 
relevant gauge group.

Now we are in a position to define our disorder operators $\{M(x) \}$. $M(x_0)$ is an operator of large gauge transformation which
acts on fields with a gauge transformation $U^(x)$
which is singular at the point $x_0$.
Consequently this transformation has
the property that as $x$ encircles $x_0$, $U$ does not return to its original value but acquires a $Z_N$ phase:
\beq
U(\phi=2\pi)=e^{-\frac{i2\pi k}{N}} U(\phi=0)
\label{U}
\eeq

From this definition we can easily find the expression for $\la M\ra$
in our case. Under a large gauge transformation the expression for the partition function of the gauge theory in the non-trivial
vacuum with label $k$ (\ref{bloop}) acquires a phase
\bea
\la k|M|k\ra  = & \frac{1}{Z} 
\int {\cal D} \psi {\cal D} \bar{\psi} {\cal D}A \left[
\exp{(- \int d^2 x ~\lie )}~
\tr_R P \exp{(i \oint_{C \rightarrow \infty} dx^\mu A_\mu (x) )} \right. \label{M} \\ 
& \left. \tr_R P \exp{(i \oint_{C \rightarrow \infty} dx^\mu U^{\dagger}i
\partial_{\mu} U  )}  \right] = e^{-\frac{i2\pi k}{N}} 
~~~,~~~ k=0,1,...N-1 \nonumber
\eea
 where $U$ is the matrix of the large gauge transformation   (\ref{U}).
Therefore, the vacuum expectation  value
of the disorder operator is a constant - a condensate
of $M$ in some sense - and its phase is a label of
the specific vacuum state in which we are.
Notice that this situation is very similar 
to what we have in a two dimensional Abelian $\theta$ vacuum with the only difference being that the label which 
marks the vacuum state in our case is a discrete rather than a
continuous variable ($\theta \in [0,2 \pi]$).  Hence we see that 
the disorder operator $M$ is a number operator for the 
vacua, returning the associated phase when acting on a particular
vacuum state.

Two remarks are in order. First of all, operator $M$ as it is
defined is not gauge invariant. It can be easily seen
from the formula (\ref{U}) there is an explicit dependence on
the position of singularity $x_0$. This phenomenon is similar
to the analysis of the Dirac string attached to a magnetic monopole where the position of the string depends on the gauge but the
fact of its existence does not. In our calculation of the vacuum
expectation value of $M$, we integrate over all of 1+1 dimensional
space-time and thus the contribution of the string is recorded.
Therefore $\la M\ra$ is a gauge invariant characteristic 
even though the operator $M$ itself is not.

Our second remark concerns the formula (\ref{M}).
As we discussed earlier, the external Wilson loop is in a fundamental
representation therefore the gauge transformation defined by
the formula
(\ref{U}) leads to the nontrivial phase in the calculation
of $\la M \ra$ (\ref{M}). If we had considered an adjoint external
loop instead we would have got the trivial result $\la M\ra=1$.
Such a result certainly means that any  adjoint external loop
does not describe a new vacuum state, but rather, is equivalent to
the trivial one - a fact which we already knew.  However, what is 
useful from this point of view is we see the importance of the center
$Z$ of the gauge group directly. Only elements nontrivial
with respect to the center may lead to the nontrivial phases
in (\ref{M}).

Because (\ref{M}) is so important we would like to explain this result
in another way. From an analysis of the 't Hooft commutation
relations \cite{Hoof1} it follows
\beq
\label{Hoof}
W(C)M(x)=M(x)W(C)\exp(\frac{2i\pi n}{N})
\eeq
Here $W(C)$ is Wilson loop operator and $n$ counts the number
of times that $C$ winds around $x$. It is easy to see from  the
original expression (\ref{bloop}) that an extra
Wilson loop insertion (one above another) generates
the following transformation of the vacuum  label $k$: 
$ k\rightarrow k+1$.
So, the Wilson operator acts on $k$-variable as a ladder operator
and takes us from the $k$ vacuum to the $k+1$ vacuum state.
Of course we have all seen this before in the standard
commutation relations of quantum mechanics: If the operator $P=-i\frac{d}{dx}$ is momentum operator, then the unitary 
operator for a finite translation $a$ 
is $\exp (iaP)$. Taking into account the commutation
relation $[P,x]=-i$, we have the identity
\beq
\label{qm}
e^{iaP}e^{ix}=e^{ia}\cdot e^{ix}e^{iaP},
\eeq
which is analogous to the 't Hooft relation (\ref{Hoof}) with the obvious substitution:
\beq
e^{iaP}\rightarrow W ;~~~~~e^{ix}\rightarrow M;~~~~~e^{ia}\rightarrow
e^{\frac{i2\pi}{N}}.
\eeq
From this analogy we see one  more time that $\la M\ra$ should have the
exponential form (\ref{M}) with a phase determined by the amount of finite
translation.

Finally we note that such a behaviour of
vacuum expectation value (\ref{M}) is in a perfect agreement with
't Hooft's conjecture about properties of the  vacuum condensation
of the dual variable in confinement phase.
In our case we can go even further though because the absence of
propagating degrees of 
freedom makes possible the calculation of 
not only $\la M\ra$ itself, but
the correlation functions like $\la M(x_1), M^{\dagger}(x_2)\ra=\la M \ra\cdot\la M^{\dagger} \ra\sim const$ as well. They simply factorize.
In general case, when physical degrees of freedom are present
the similar behaviour remains the same  only in the limit
$(x_1-x_2)^2\rightarrow\infty$ as a manifestation
of the standard properties of cluster decomposition.

\section{Conclusion}

We have explicitly calculated string tensions between heavy adjoint
charges and consequently found the number of vacuum states
in two dimensional gluodynamics for each compact Lie
symmetry group. It was found that string tension, and consequently
the spectrum of the theory, depends on which vacua one is in.
Additionally, it was shown that the number of vacuum states 
is in agreement with topological classification and it is given by
$\pi_1$ of the corresponding effective gauge 
group as documented in Table 1.
We believe that
the number of absolutely stable vacua will remain the same even if one
introduces the physical degrees of freedom so long as the system 
respects the center symmetry.
The  most important   observation which can be derived from this study
is the existence of a new superselection rule which must be imposed
in order to count these new vacuum states.
One could always say that an introduction of fundamental
fields will destroy this classification: therefore
there is no reason to study configurations which
can not survive in the presence of fundamental matter.
It is well known, however, why this argument is flawed.
It is believed  that the main configurations in $QCD_4$
responsible for the confinement are pure gluon (not quark)
configurations and, at least in the large $N_c$ limit this statement
is certainly correct. Therefore, one could expect
that an introduction of fundamental fermions might change
some qualitative characteristics, but not a quantitative picture
derived in pure gluodynamics.

We believe that this is also true in $QCD_2$ where we essentially
introduced and studied this new superselection rule.
We hope that the future investigation  of these,
fundamentally new superselection rules and  
vacuum states may shed the light on the many issues
raised in the Introduction regarding supersymmetric models,
chiral condensates and behaviour in the large $N_c$ limit.
We feel that adopting a point of view about classifying 
vacuum states as we have outlined here may lead to progress
in resolving current dilemmas.

Finally, it might happen that a useful tool to formulate
(and answer) questions about vacuum structure is a disorder 
operator in the dual description.
Rather than the Wilson loop operator it is the disorder operator
which plays the fundamental role of enumerating the  
vacua in the theory.
At least recent developments \cite{seiberg} in supersymmetric models
are based on the dual picture. In this approach the operator of the
creation of a point monopole can be regarded as a large gauge
transformation with the corresponding  properties.
This operator is in a sense  analogous to our disorder operator (\ref{M}) and, as is known, the monopole condensation (which
is formally described as a nonzero condensate of the
monopole creation operator) is the fundamental property of the 
theory with confinement. Of course, our operator (\ref{M})
is in some sense trivial because it does not create any physical
degrees of freedom. However, we believe that the main topological
properties of this operator remain the same even in the
presence of physical degrees of freedom.

\section*{Acknowledgment} 

We thank Sasha Polyakov for discussions and for the historical remarks
regarding the disorder operator in the  dual  description.
This work is supported in part by the Natural Sciences and 
Engineering Research Council of Canada.  L.P. is supported in 
part by a University of British Columbia Graduate Fellowship.

\section{Appendix}

Here we explain some of the notations and details of the group 
theory used in our analysis.  First we need to define the Young tableaux
notation we use.  Since any tensor representation of a rank n 
compact Lie group  may be denoted by a Young table we will deal with
convenient characterizations of these tables.  Let $l_i$ be the number
of boxes in the $i$th row of the Young table for some
representation $R$ then  then we have non-increasing sequence
$l_1 \geq l_2 \geq \ldots \geq l_n $.
In the text this sequence is denoted $(l_1  l_2 \ldots l_n)$.

Expressions for the quadratic Casimirs of  tensor representations
for the classical are given in terms of simple formulae \cite{them}
of the table variables ${l_i}$. If we denote the total number of 
boxes in the Young table for representation $R$ by $l=\sum l_i$ then
\beq
SU(N) : ~~~~~~~C_2(R)=  \sum_{i=1}^N l_i (N + l_i +1 -2 i -l/N)
\eeq
\beq
Sp(2N) : ~~~~~~~C_2(R)= \frac{1}{2} \sum_{i=1}^N l_i (2N + l_i +2 -2 i)
\eeq
\beq
SO(N) : ~~~~~~~C_2(R)=  \sum_{i=1}^N l_i (N + l_i -2 i)
\eeq

Of particular interest for us are the details of adjoint representations 
for different groups. These are tensor representations for every compact
Lie group and, hence, can be stated in terms of table variables $l_i$.
For the classical Lie groups we have the following table configurations
for adjoint representations
\beq
SU(N) : ~~~~~~~l_1=2,~l_2 \ldots \l_{N-1} =1 
\eeq
\beq
Sp(2N) : ~~~~~~~ l_1=2,~l_2 \ldots \l_N =0
\eeq
\beq
SO(2N+1) : ~~~~~~~ l_1=l_2=1,~l_3 \ldots \l_N=0
\eeq
\beq
SO(2N) : ~~~~~~~ l_1=l_2=1,~l_3 \ldots \l_N=0
\eeq 

For representations of the exceptional groups and spinor representations
of the orthogonal groups one may calculate quadratic Casimirs directly 
from knowledge of the weights and the Weyl formula but for our
purposes it was sufficient to  use existing tables \cite{branch} and 
\cite{slansky}.

\newpage
 
\epsffile[32 105 390 660]{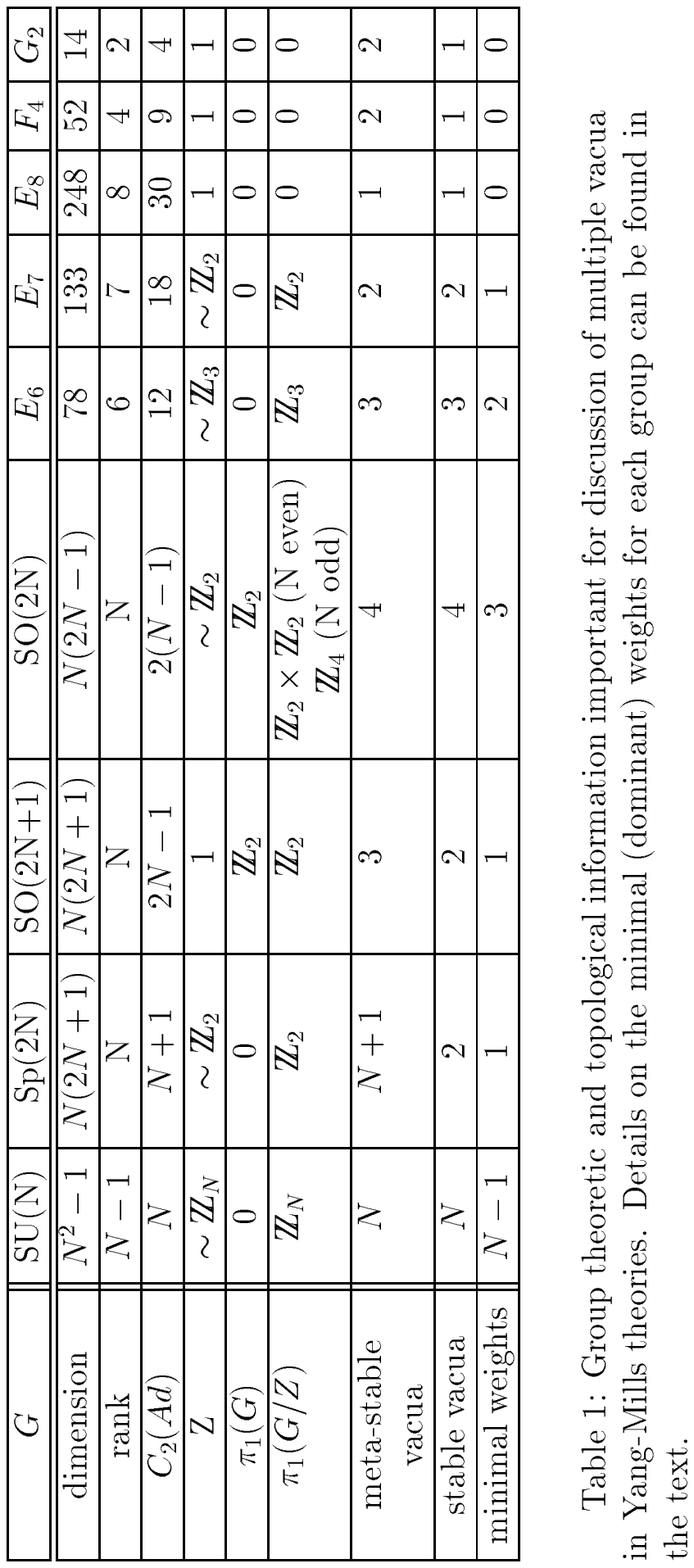}
 
\end{document}